\documentclass{elsart}
\usepackage{graphicx}
\usepackage{amsmath}
\usepackage{amssymb}
\usepackage{epsfig}
\newcommand{\ket}[1]{\lvert#1\rangle}

\begin{document}

\begin{frontmatter}

\title{Spin excitations in the Fractional Quantum Hall regime at $\nu\lesssim1/3$}

\author[address1]{Y. Gallais},
\author[address1]{T.H. Kirschenmann},
\author[address1]{C.F. Hirjibehedin},
\author[address1]{I. Dujovne},
\author[address1,address2]{A. Pinczuk}
\author[address2]{L.N. Pfeiffer} and
\author[address2]{K.W. West}

\address[address1]{Departments of Physics and of Applied Physics, Columbia University, New York, NY 10027, USA}

\address[address2]{Bell Labs, Lucent Technologies, Murray Hill, New Jersey 07974, USA}

\thanks[thank1]{
Corresponding author.
E-mail: yann@phys.columbia.edu}

\begin{abstract}
We report inelastic light scattering experiments in the fractional
quantum Hall regime at filling factors $\nu\lesssim1/3$.  A spin
mode is observed below the Zeeman energy. The filling factor
dependence of the mode energy is consistent with its assignment to
spin flip excitations of composite fermions with four attached flux
quanta ($\phi$=4). Our findings reveal a composite fermion Landau
level structure in the $\phi$=4 sequence.
\end{abstract}

\end{frontmatter}

\section{Introduction}
In the composite fermion (CF) picture of the fractional quantum Hall
(FQH) effect, fundamental interactions are taken into account at the
mean field level by mapping the system of strongly interacting 2D
electrons in magnetic field into a system of weakly interacting
Composite Fermions (CF) moving in a reduced effective magnetic field
\cite{Jain}. The reduction in magnetic field follows from the
binding of $\phi$ flux quanta to electrons, so that effective
magnetic field experienced by CF quasiparticles is $B^*=\pm B/(\phi
p\pm 1)$, where $p$ is an integer that enumerates members of a
particular sequence and $\phi$ is a even integer that labels
different sequences. In this picture, the FQH effect can be
understood by the emergence of CF Landau levels  with cyclotron
frequency: $\omega_{CF}=\frac{eB^*}{cm^{*}}$, where $m^{*}$ is an
effective CF mass. Evidence for a spin split Landau level structure
of CF for the $\phi=2$ sequence has been provided by
magnetotransport experiments in tilted magnetic fields at filling
factors near $\nu=3/2$ \cite{Du} and by inelastic light scattering
studies of spin excitations in the range $1/3<\nu<2/5$ \cite{Irene}.
For the $\phi=4$ sequence  (i.e. $\nu\lesssim1/3$), however, direct
evidence for such CF Landau level structure is lacking. Studies of
the $\phi=4$ sequence are more difficult because of the higher
magnetic fields that are required and, compared to the $\phi=2$
sequence, the smaller energy scales in the excitations. Insight on
the energy scales for excitations were revealed by activated
transport measurements \cite{Pan} and by the recent observations of
$\phi=4$ quasiparticle excitations in light scattering experiments
\cite{Cyrus}.

In this work, we present a resonant inelastic light scattering study
of spin excitations for $\nu\lesssim1/3$. The excitations are spin
waves (SW) and spin-flip (SF) modes. The SF excitations involve a
change in both the spin orientation and CF Landau level quantum
number. We monitor the evolution of these spin excitations below and
away from $\nu=1/3$ when the population of the excited CF Landau
level increases. Our results reveal the existence of spin split CF
Landau levels in the $\phi=4$ sequence. The SW-SF splitting is
linear in magnetic field. This determination suggests an effective
mass significantly larger than the activation mass of CF with
$\phi=4$.
\section{Sample and Experiment}
The 2D electron (2DES) system studied here is a GaAs single quantum
well of width $w=330~\AA$. The electron density at small magnetic
fields is n=5.5$\times$10$^{10}$~cm$^{-2}$ and the low temperature
mobility is $\mu$=7.2$\times$10$^6$/Vs. The sample is mounted in a
backscattering geometry, making an angle $\theta$ between the
incident photons and the normal to the sample surface. The magnetic
field perpendicular to the sample is B=B$_T$cos$\theta$, where B$_T$ is
the total applied field. The results reported here have been
obtained at $\theta$=50$\pm$2 degrees. Similar results have been seen at 30 degrees \cite{Cyrus-thesis}.The ingoing and outgoing
polarizations were chosen to be orthogonal (depolarized spectra)
since excitations which involve a change in the spin degrees of
freedom are stronger in this configuration. The sample was cooled in
a dilution refrigerator with windows for optical access. All the
measurements were performed at the base temperature T=23~mK and the
power density was kept below 10$^{-5}$W/cm$^2$ to avoid heating the
electron system. The energy of the incident photons was tuned to be
in resonance with the excitonic optical transitions of the 2DES in
the FQH regime \cite{Goldberg,Bar-joseph,Cyrus2}.
\section{Results and discussion}
\begin{figure}
\centering \epsfig{figure=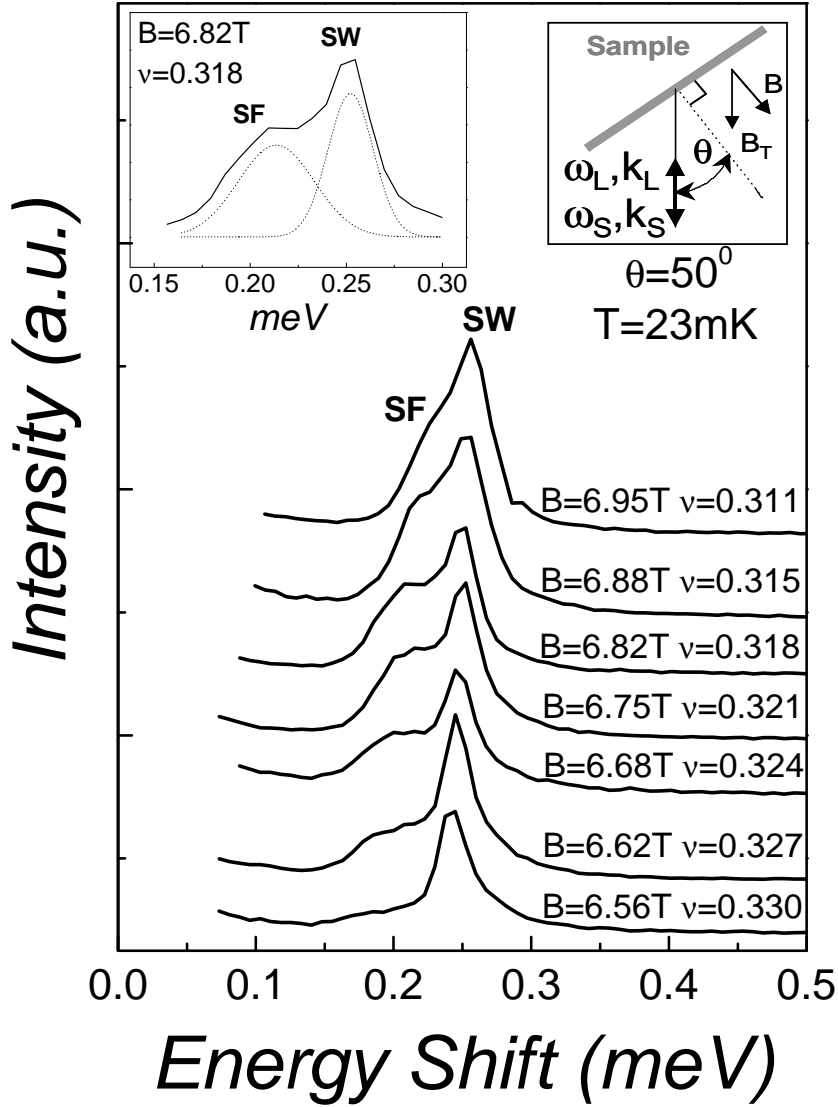, width=0.79\linewidth, clip=}
\caption{Low energy spectra of spin excitations in the filling
factor range 0.31$<\nu<$0.33. The most intense peak is the long
wavelength spin-wave at the Zeeman energy E$_z$ while the peak its
low energy side is assigned to a spin-flip transition (see text and
figure \ref{levels}). The left inset shows the result of a
two-gaussian fitting procedure for the two peaks to extract their
respective energies. The right inset shows the backscattering
configuration} \label{spectres}
\end{figure}
\begin{figure}
\centering \epsfig{figure=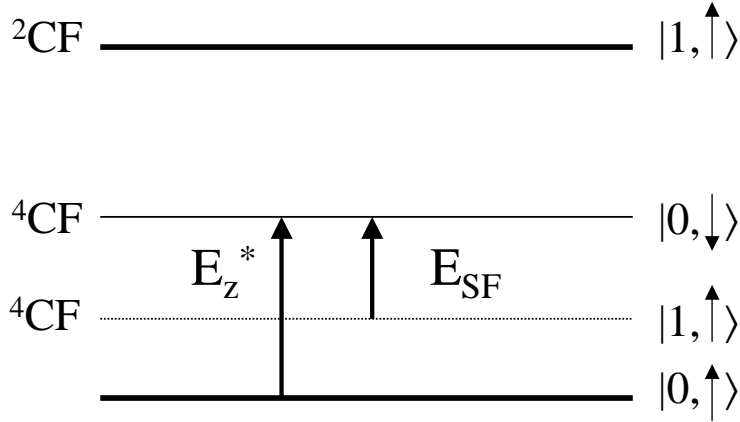, width=0.7\linewidth, clip=}
\caption{Structure of spin split CF Landau levels for 2/7$<\nu<$1/3 ($\phi=4$).
Two $^4$CF spin transitions are possible. The large $q$ spin wave is
at E$^*_z$=E$_z$+E$^{\uparrow\downarrow}$ where
E$^{\uparrow\downarrow}$ is the spin reversal energy. The spin-flip
excitation at E$_{SF}$. The spin-flip excitation emerges when the
$\ket{1,\uparrow}$ level is populated.} \label{levels}
\end{figure}

Figure \ref{spectres} shows the evolution of the low energy spectrum
for $\nu\lesssim1/3$. $\nu=1/3$ corresponds to a perpendicular field
of 6.5T and the filling factor range studied corresponds to the
range 0.31$<\nu<$0.33. Close to $\nu=1/3$, the spectra are dominated
by the long wavelength SW at the 'bare' Zeeman energy
E$_z=g\mu_BB_T$, where g=0.44 is the Lande factor for electrons in
GaAs and $\mu_B$ is the Bohr magneton. For filling factors away from
$\nu=1/3$ an excitation emerges on the low energy side of E$_z$. We
assign this excitation to a SF mode linked to transitions in the CF
framework that involve the first excited CF Landau level as depicted
in figure \ref{levels} . At $\nu=1/3$ the first CF Landau
($\ket{0,\uparrow}$) level is fully occupied while for $\nu=2/7$ the
first two CF Landau levels ($\ket{0,\uparrow}$ and
$\ket{1,\uparrow}$) are occupied. In between the two incompressible
states, the first excited Landau level is partially populated and SF
transitions between $\ket{1,\uparrow}$ and $\ket{0,\downarrow}$
starting from the partially filled level become possible. Thus the
study of the SF excitations in the filling factor range
2/7$<\nu<$1/3 probe directly the CF level structure for $\phi=4$.
For small occupation of the $\ket{1,\uparrow}$ excited level and
when the coupling between the excited quasiparticle and its
quasihole is negligible, the SF transition energy can be written as
in the $\phi=2$ case:
\begin{equation}
E_{SF}=E_z+E^{\uparrow\downarrow}-\hbar\omega_c
\label{ESF}
\end{equation}
where E$^{\uparrow\downarrow}$ is the spin reversal energy which is
a measure of the residual interactions between $\phi=4$ CF
quasiparticles \cite{Pinczuk,Longo,Aoki,Mandal,Irene}.
\begin{figure}
\centering \epsfig{figure=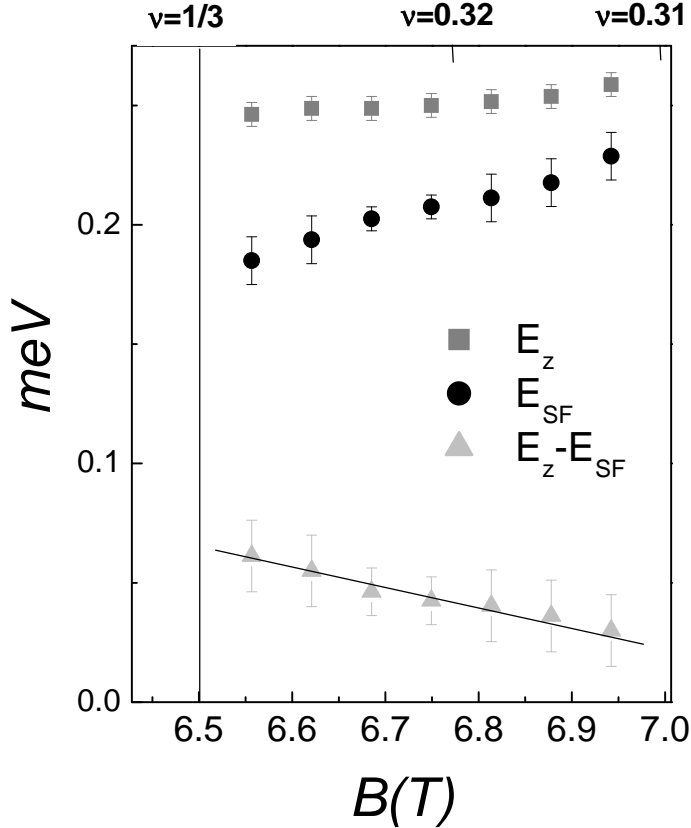, width=0.65\linewidth, clip=}
\caption{Magnetic field dependence of the Zeeman (E$_z$) and
spin-flip excitation energies for 0.31$<\nu<$0.33. Also shown is the
evolution of the splitting E$_z$-E$_{SF}$.} \label{energies}
\end{figure}
The energy E$_{SF}$ was extracted for each filling factor by
performing a simple analysis of the low energy spectra using a
two-gaussian fitting procedure as shown in the inset of figure
\ref{spectres}. Figure \ref{energies} displays the corresponding
energies, E$_z$ and E$_{SF}$ as a function of filling factor. The
strong dependence of E$_{SF}$ confirms our assignment of the peak as
excitation involving spin degrees of freedom. More importantly, the
spacing between E$_z$ and E$_{SF}$ is not constant and decreases
with the filling factor. From equation \ref{ESF}, we easily see that
this spacing is directly related to the CF cyclotron energy so that
the splitting between the two spin excitations is
E$_z$-E$_{SF}$=$\hbar\omega_c$-E$^{\uparrow\downarrow}$.
The magnetic field dependence of the E$_z$-E$_{SF}$ spacing is set
by the effective field B$^*$. For the $\phi=2$ sequence, $^2$CF
emanate from the $\nu=1/2$ state and the effective field has its
origin at B$_{1/2}$. For the $\phi=4$ sequence however, $^4$CF
emanate from the $\nu$=1/4 state and the origin is at B$_{1/4}$. and
the effective magnetic field should then decrease when going from
0.33 to 0.31. This is indeed consistent with our data and to the
existence of $^4$CF or $\phi=4$ spin-flip excitations below
$\nu$=1/3. Our results support the CF Landau level picture shown figure
\ref{levels} for the $\phi=4$ sequence.

The linear decrease of E$_z$-E$_{SF}$ with the effective magnetic
field makes very tempting the evaluation of an effective mass by using
a slope that is simply given by $\frac{\hbar e}{m^{*}c}$ in the
framework of equation \ref{ESF} for E$_{SF}$. Our data between
filling factors 0.33 and 0.31 give m$^{*}$=1.5($\pm 0.1$)~m$_e$
where m$_e$ is the bare electron mass. Previous determinations using
the activation gap values at $\nu$=2/7, 3/11 and 4/15 on samples
with similar densities yield values around 0.9~m$_e$ \cite{Pan}. We
note that while our determination of m$^{*}$ is performed between
$\nu$=0.33 and $\nu$=0.31, i.e. in between the incompressible states
at 1/3 and 2/7, the effective mass determined via transport
measurements comes from a linear scaling of the activation gap
values at the incompressible states. The high
effective mass obtained in our analysis may be linked to the onset
of significant CF interactions in the partially populated CF Landau
level.
\begin{figure}
\centering
\epsfig{figure=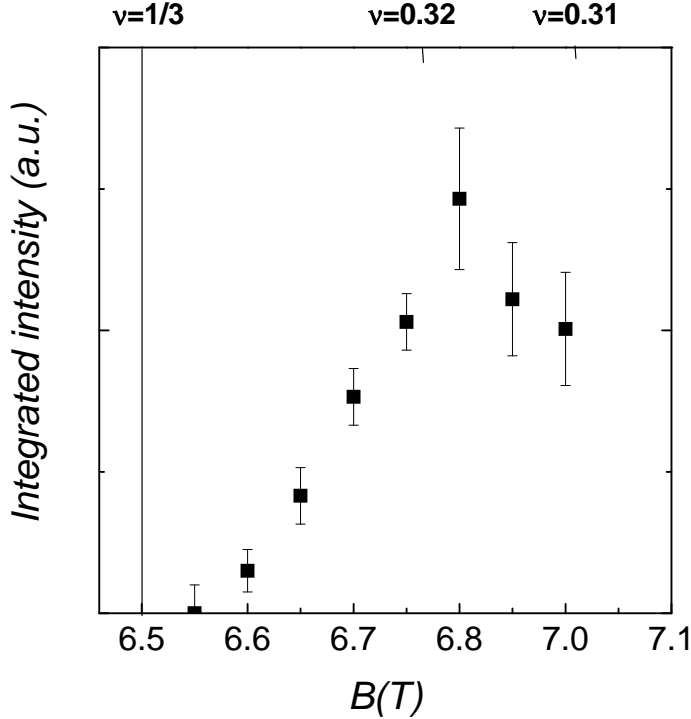, width=0.65\linewidth, clip=}
\caption{Evolution of the SF integrated intensity for 0.31$<\nu<$0.33.}
\label{intensities}
\end{figure}
Additional insights can be obtained by tracking the evolution of the
intensity of the SF excitation when the population of the first
excited CF level increases. This is done in Fig. \ref{intensities}
where the integrated spectral weight of the SF excitation is plotted
as a function of filling factor. As expected, the SF intensity
increases when the population of $\ket{1,\uparrow}$ increases but
displays an intriguing saturation around below $\nu$=0.32. As
already mentioned for the effective mass, the saturation may result
from increasing impact of CF residual interactions. These
interactions could possibly lead to further condensation into higher
order CF in the partially populated level. Recent transport
measurements have indeed shown the possible existence of such higher
order states even for the $\phi$=4 sequence \cite{Pan2}.
\section{Conclusion}
In this study, we have shown the existence of spin-flip excitations
of $^4$CF quasiparticle below $\nu$=1/3. The results indicate the
existence of a spin-split CF Landau level structure for the $\phi=4$
sequence of the fractional quantum Hall effect that is similar to
the one found for the $\phi=2$ sequence. The evolution of the SF
energy with effective magnetic field yields an effective mass
of 1.5m$_e$. The evolution of the SF intensity with filling factor
might signal the onset of significant CF-CF interactions that could
possibly lead to further CF condensation.

This work is supported by the National Science Foundation under Grant No. NMR-0352738, by the Department of Energy under Grant No. DE-AIO2--04ER46133, and by a research grant from the W.M. Keck Foundation.

\end{document}